\documentclass[reprint,amsmath,amssymb, aps,prb,superscriptaddress,nofootinbib,twocolumn]{revtex4-2}
\usepackage{subcaption}
\usepackage[utf8]{inputenc}
\usepackage{braket}
\usepackage{graphicx}
\usepackage{lipsum}
\usepackage{mathtools}
\usepackage{standalone}
\usepackage{cuted}
\usepackage{xcolor}
\usepackage{bm}
\usepackage{hyperref}
\usepackage[mathlines]{lineno}
\usepackage{wrapfig}
\usepackage{siunitx}

\DeclareSIUnit\photon{photon}
\usepackage{dcolumn}
\usepackage{booktabs}
\usepackage{tabularx} 
\usepackage{tikz}
\usetikzlibrary{fit}
\usepackage{pgfplots} 
\usepgfplotslibrary{fillbetween} 
\definecolor{tensorblue}{rgb}{0.8,0.8,1}
\tikzstyle{tensor}=[rectangle, draw=black, fill=tensorblue, thick, minimum size = 6mm]

\pgfplotsset{compat=1.18}
\usepackage{hyperref}
\hypersetup{
    colorlinks,
    citecolor=black,
    filecolor=black,
    linkcolor=black,
    urlcolor=black
}
\usepackage{subcaption}

\DeclareMathOperator{\Tr}{Tr} 
\usepackage{bbold}
\tikzstyle{mynode}=[thick,draw=blue,fill=blue!20,minimum size=22] 

\usetikzlibrary{quantikz}
\usepackage[normalem]{ulem}

\date{\today}
\begin{document}
\title{High Capacity Noisy Unruh--DeWitt Quantum Channels with Bosonic Dephasing}

\author{Eric W. Aspling}
\affiliation{Department of Physics, Applied Physics, and Astronomy, Binghamton University, Binghamton, NY 13902}
\author{Michael J. Lawler}
\affiliation{Department of Physics, Applied Physics, and Astronomy, Binghamton University, Binghamton, NY 13902}
\affiliation{Department of Physics, Cornell University, Ithaca, NY 14853}
\affiliation{Department of Physics, Harvard University, Cambridge, MA 02138}
\begin{abstract}
Unruh--DeWitt (UDW) detectors implemented as UDW quantum gates provide a framework for evaluating quantum Shannon theory properties of qubit-field systems. UDW quantum channels consist of qubits encoding/decoding quantum information onto/off of quantum fields. With the controlled unitary structure of UDW gates, the encoding/decoding process happens on the diagonals of the coherent state density matrix describing the field. However, given the non-orthogonality of coherent states the output of UDW channels consists of unwanted states and unwanted mixing of states that lower the channel capacity. In idealized models, these appear in the off-diagonals and diagonals of the field's density matrix in the coherent state basis. For this reason, we show that UDW quantum channels have an unexpected representation as certain bosonic dephasing channels with dephasing parameters captured by a combination of the coupling, smearing, and switching functions of the UDW detector model. We demonstrate the unexpected consequence that a larger dephasing parameter results in higher channel capacity and helps alleviate unwanted state mixing. We illustrate these properties through two examples: inserting an additional ideal dephasing channel into the quantum channel and inserting cross-talk noise via a third UDW gate. Remarkably, the cross-talk noise channel qualitatively improves a lower bound on the quantum capacity suggesting UDW gates will have unexpected performance improvements if realized in condensed matter experiments. 
\end{abstract}
\maketitle
\section{Introduction}
Relativistic Quantum Information (RQI) has benefited tremendously by introducing Unruh--DeWitt (UDW) detectors modeling qubit-field interactions. Among these benefits includes introducing RQI to quantum Shannon theory \cite{Martin-Martinez2011relativistic,Simidzija2020Transmission,Sachs2017Entanglement,Simidzija2018General,Yamaguchi2023Entanglement,Tjoa2022Channel}. Often, quantum channels that utilize UDW gates involve encoding (decoding) quantum information onto (off of) fields using coherent states to represent the energy levels of the field. The non-orthogonality of coherent states can lead to information scrambling when the field is traced out as unwanted states carry non-zero weights. Conveniently, the parameters of the UDW model can be constrained to provide experimentally plausible quantum computers \cite{aspling2022design}. One particular constraint, that of strong coupling, is reminiscent of the role the dephasing parameter has in bosonic dephasing channels. As we will see, these are the same.  

Encoding and decoding information onto and off of fields has been theoretically demonstrated with great success over the past decade. The coupling, smearing, and switching functions of the UDW model provide the necessary parameters to fine-tune the interactions between the qubits and fields, allowing for near-perfect quantum capacity in limits of strong coupling \cite{Simidzija2020Transmission,aspling2022design}. The ability to demonstrate other quantities of Shannon theory, such as the impact of noise, will further solidify the current and future role UDW gates play in quantum computing and condensed matter systems.

Dephasing channels \cite{Wilde2011From} represent the effects of noise on the off-diagonals of any input density matrix. Dephasing channels can be used to model decoherence as well as other effects that realize this prescription. Bosonic dephasing channels \cite{lami2023exact,Arqand2020Quantum,dehdashti2022quantum,Jiang2010Evaluating,Leviant2022quantum} have been the center of an open problem for more than a decade \cite{Jiang2010Evaluating} and now with the recent advances in RQI, can be used to model decoherence of quantum channels between qubits and fields. The key connection here is the usage of coherent states to model the excitations of fields in the UDW quantum gate model.  

Coherent states \cite{kam2023coherent} are powerful tools for understanding bosonic channels, and in the setting of UDW gates, they provide the encoded states for carrying out state-transfer between qubits and fields \cite{Simidzija2018General,Simidzija2020Transmission,Tjoa2022Channel}. Given the non-orthogonality of coherent states, the coupling, smearing, and switching parameters are designed to constrain the inner products of the coherent states. This is a matter of consequence for tracing over the field after the quantum information has been transmitted from one qubit to another (via. the field). In the weak coupling regime, these parameters provide a channel capacity near zero. \textbf{For this reason, we require strong coupling where the unwanted terms go to zero in the limit of infinite coupling strength.} 

Unruh--DeWitt quantum computing \cite{aspling2022design} is a new field that aims to understand entanglement propagation in quantum materials. As with any quantum computing setup, modeling noise is a key component of the theory. Understanding UDW channels from a bosonic dephasing channel perspective not only indicates where the noise comes from but provides an avenue for numerically modeling noise and subsequently offers additional solutions.

Therefore, we find several areas of these parameters worth exploring in the context of bosonic dephasing channels. Firstly, the current method of fine-tuned parameters fits nicely into the context of a ``dephasing rate". The constant associated with the dephasing rate, commonly denoted as $\gamma$ is some prefactor of the coherent amplitude. As we will show, this prefactor exists in the coupling, smearing, and switching functions of the UDW detector model. Adjustments to these parameters explicitly change the coherent information of a given qubit-field UDW channel, as one would expect of any dephasing rate constant. Secondly, with this new perspective in place, noise via. similar interactions as well as interactions with the environment, should be calculable quantities. We show that in the context of the idealized canonical bosonic dephasing channel, one where the system interacts with the environment, no such change to the quantum information is present. This lack of effect is a consequence of the input already being constrained to near-zero valued off-diagonals, so any additional constraints to the off-diagonals are negligible. Lastly, we demonstrate with a given probability distribution the ability to calculate the coherence of a noisy UDW quantum channel, where the noise is a consequence of some addition of cross-talk noise that follows from unwanted interactions with unknown detectors.

\section{UDW Quantum Channels}\label{background}
The overlap between the communities of quantum Shannon theory and RQI is minimal. Namely, the overlap exists mostly as those working directly with UDW detectors in RQI. For those familiar with this setup, feel free to skip to Sec.~\ref{Unruh--DeWitt Dephasing Channels}. For the rest, we take this section to outline the setup of the UDW detector channel and explain the constraints that will be utilized when recasting our UDW channel from a dephasing perspective.  
\subsection{Coherent States} \label{Coherent States}
A 1-D scalar field $\hat{\varphi}(x)$ and it's associated conjugate momentum $\hat{\Pi}(x)$ are given as
\begin{flalign}
     \hat{\varphi}(x) &= \int\frac{dk}{2\pi} \sqrt{\frac{v}{2\omega(k)}}[\hat{a}(k)e^{ikx}+\hat{a}^{\dagger}(k)e^{-ikx}] \label{scalar field}\\
     \hat{\Pi}(x) &=\int\frac{dk}{2\pi} \sqrt{\frac{\omega(k)}{2v}}[-i\hat{a}(k)e^{ikx}+i\hat{a}^{\dagger}(k)e^{-ikx}].\label{conjugate momentum}
\end{flalign}
In this continuous space, these field observables can be expressed as displacement operators  $D[\alpha(x)]$ with a point-wise coherent amplitude
\begin{equation}\label{real space coherent amplitude}
    \alpha_{\varphi}(x) = \sqrt{\frac{v}{2\omega(k)}}e^{-ikx}.
\end{equation} 
For convenience we have introduced subscripts to differentiate between coherent amplitudes of the  field and conjugate momentum observables. A Fourier transform following the usual prescription
\begin{equation}
f(k) \coloneqq \frac{1}{\sqrt{(2\pi)}}\int dx f(x) e^{ikx}
\end{equation}
allows us to express a continuous displacement operator of a scalar bosonic field\cite{Simidzija2017Nonperturbative} as 
\begin{multline}\label{field coherent state definition}
    \exp{(\pm i \hat{\varphi})}\ket{0} = \hat{D}[\alpha(k)]\ket{0}\\
    = \exp{\left(\int dk\left[\alpha(k)\hat{a}_{k}^{\dagger}-\alpha(k)^*\hat{a}_{k}\right]\right)} \ket{0} \equiv \ket{\pm \alpha}.
\end{multline}
The final equivalence is a notational convenience we will adopt in Sec.~\ref{Two Applications of a UDW Dephasing Channel} when carrying out simulations. This notational convenience can be understood as single mode excitation of a continuous spectrum. For now we proceed with the more general form.

These field observables follow the equal-time canonical commutation relations $[\hat{\varphi}(x),\hat{\varphi}(x')]=0 ,\, [\hat{\Pi}(x),\hat{\Pi}(x')]=0,$ and $[\hat{\varphi}(x),\hat{\Pi}(x')] = i \delta(x-x')$ and consequently produce displacement operators that obey commutation relations 
\begin{equation} \label{Displacement commutation relation}
    [\hat{a}_{\varphi}(k'), \hat{D}(\alpha_{\varphi}(k))] = \alpha_{\varphi}(k')\hat{D}_{\varphi}(\alpha(k)).
\end{equation}

\subsection{Unruh--DeWitt Quantum Gates} \label{Unruh--DeWitt Quantum Gates}
A unitary operator in the form of a quantum gate can generate and/or propagate quantum information through a quantum circuit \cite{nielsen2010Quantum}. We expect this same consequence to be achievable with UDW detector models \cite{Simidzija2020Transmission}. Incorporating our scalar field from Eq.~\ref{scalar field} we present the well known UDW detector model as 
\begin{equation}\label{time-ordering UDW detector}
U_{UDW}(t_1,t_2) = Te^{-i\int_{t_1}^{t_2}dt \hat{\mathrm{H}}_{int}(t)}
\end{equation}
with
\begin{equation}\label{UDW model}
    \hat{\mathrm{H}}_{int}(t) =  \lambda\chi(t) \int_{\mathbb{R}} dx \ F(k) \hat{\mu}(t) \otimes \hat{\varphi}(k,t).
\end{equation}
where $\chi(t)$ and $F(k)$ are the switching and smearing functions respectively and $\lambda$ is a time dependent coupling constant that indicates what gate is interacting at what time and $\hat{\mu}(t)$ is a two state (qubit) detector that has the form
\begin{equation}
    \lambda\hat{\mu}(t) = \frac{\lambda}{2}(\hat{S_+} e^{-i\Omega t} + \hat{S_-}e^{+i\Omega t}) \equiv \lambda \hat{\mu}(t)
\end{equation}
where $\hat{S}_i$ is the projector onto some spin state and $\Omega$ is some real valued number that indicates a non-trivial energy difference between spin states. By using a delta like switching function we can indicate the time ordering of our Unitary and find that is simplifies to
\begin{equation} \label{SimpleUnitary 1}
    \hat{U}_{UDW} = \exp{(-i\lambda\hat{\mu}\otimes \hat{\varphi})}.
\end{equation}

The structure of $\hat{\mu}$ as discussed above is that of a qubit observable and for this we can rewrite the gate by introducing projectors $\hat P_s$ onto the eigenstates of $\hat\mu$ as
\begin{equation} \label{SimpleUnitary projector and field}
    \hat{U}_{UDW} = \sum_{s \in \pm}\hat{P}_s\otimes e^{is\hat{\varphi}(F)}.
\end{equation}
In this form we see that $\hat U_{UDW}$ is like a controlled gate, it acts with one unitary on the field in one eigenspace of the qubit and another unitary in the other eigenspace of the qubit. This is part of the ``encoding" process that imparts the quantum information from the qubit to the field.
Here we have redefined the field observables to include the smearing and coupling of the UDW model
\begin{flalign}
    \hat{\varphi}(F) &\coloneqq \lambda_{\varphi} \int dx F(k) \hat{\varphi}(k,t)\label{redefined scalar}\\
    \hat{\Pi}(F) &\coloneqq \lambda_{\Pi} \int dx F(k) \hat{\Pi}(k,t)\label{redefined conjugate momentum}.
\end{flalign}

Equation~\ref{SimpleUnitary projector and field} captures one of the two criteria discussed in the beginning of this section, namely entanglement generation. However, given that the interest of the dephasing channel scheme is to monitor the changes in quantum capacity we instead aim for a channel that can preserves quantum information.

\subsection{UDW State Transfer Channel} \label{UDW State Transfer Channel}
\subsubsection{Setting the Channel Up}

\begin{figure}[t!]
\includegraphics{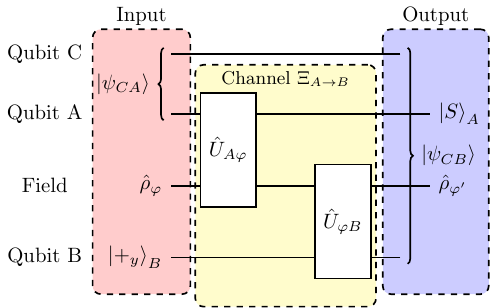}
\caption{Operators $\hat{U}_{A\varphi}$ and $\hat{U}_{\varphi B}$ are employed to encode and decode quantum information onto and off of the field $\hat{\varphi}$ thus transferring the quantum information from qubit A to qubit B.} \label{UDWQC Channel without starburst}
\end{figure}

It was shown in Ref.~\cite{Simidzija2020Transmission} that to write down a channel that preserves entanglement and performs a state transfer, we would need two of the controlled unitaries presented in Eq.~\ref{SimpleUnitary projector and field}. This will lead to the unitary
\begin{equation} \label{Simple two Unitary projector and field}
    \hat{U}_{\nu \varphi} = \sum_{z,x \in \pm}\hat{P}_x\hat{P}_z\otimes e^{ix\hat{\Pi}(F)_{\nu}}e^{iz\hat{\varphi}(F)_{\nu}}
\end{equation}
where we introduce the notation that $\hat{P}_x$, with an $x$-index, and $\hat{P}_z$, with a $z$-index, are the projection operators onto the eigenstates of Pauli matrices $\hat{\sigma}_x$ and $\hat{\sigma}_z$ respectively and $\nu$ indicates the qubit interacting with the field at a given time $t_{\nu}$. Figure~\ref{UDWQC Channel without starburst} presents a circuit diagram for this channel which has the mathematical structure 
\begin{equation}\label{UDWQC_Channel}
    \Xi_{A\rightarrow B} = \Tr_{A\varphi}[U_{A\hat{\varphi}} U_{\varphi B}( \hat{\rho}_{A,0} \otimes \hat{\rho}_{\varphi} \otimes \hat{\rho}_{B,0})U^{\dagger}_{\varphi B} U^{\dagger}_{A{\varphi}}]
\end{equation}
and can be expanded out as 
\begin{multline} \label{Expanded out Channel}
    \Xi_{A\rightarrow B}=\\ \sum_{l,l',x_i,z_i}\langle0|e^{iz_1\hat{\varphi}_A}e^{ix_1\hat{\Pi}_A}e^{ix_2\hat{\Pi}_A}e^{iz_2\hat{\varphi}_A}\\ \times e^{iz_1\hat{\varphi}_A}e^{ix_1\hat{\Pi}_A}e^{ix_2\hat{\Pi}_A}e^{iz_2\hat{\varphi}_A}|0\rangle\\ \times \braket{l'_z|\hat{P}_{-z_1}\hat{P}_{-x_1} \hat{P}_{x_4} \hat{P}_{z_4} |l_z}_C\bra{l'_z}\\ \otimes \hat{P}_{-z_3} \hat{P}_{-x_3}\ket{+_y}_B\bra{+_y}\hat{P}_{x_2} \hat{P}_{z_2}.
\end{multline}
The correlator in Eq.~\ref{Expanded out Channel}, which is a result of the trace over the field, is a tricky value to calculate. One may recognize the correlator as a vacuum expectation value of eight vertex operators, often utilized in conformal field theory and string theory. If utilizing coherent states, we recognize that they will lead to non-orthogonal inner products $|\braket{+\alpha(k)|-\alpha(k)}|$ that contain the specific unwanted states that if we could dephase away, would improve the coherent information of the channel. Furthermore, any additional interactions on the field will only yield a larger correlator. Therefore, we also explore the possibility that a dephasing perspective may eliminate the necessity to increase the size of the correlator for additional interactions.   

\subsubsection{Choosing the Correct Parameters}\label{Choosing the Correct Parameters}
Within Eq.~\ref{UDW model} and Eqs.~\ref{redefined scalar} and \ref{redefined conjugate momentum}, there are still a few free parameters remaining. Namely, the coupling constants $\lambda_{\varphi}$ and $\lambda_{\Pi}$ as well as the smearing function $F(k)$. We therefore utilize these parameters to remove unwanted final states. 

These unwanted states show up in two ways, firstly during the application of $e^{ix\hat{\Pi}}$, which when applied to a coherent state will generally change the value of the state, and secondly with non-zero values of the inner product $|\braket{+\alpha(k)|-\alpha(k)}|$. In RQI literature, it is common to redefine the coherent amplitude, given in Eq.~\ref{real space coherent amplitude}, to include the coupling constants and the smearing function. By doing this we can utilize equation \ref{Displacement commutation relation} and the following identity 
\begin{equation}
     e^{\pm i\Pi}\hat{a}_ke^{\mp i \Pi} = \hat{a}_k \mp \alpha_{\Pi}(k)
\end{equation}
to prove the relation 
\begin{equation}
    \hat{\Pi}\ket{\pm \alpha(k)} = \pm \Gamma \ket{\pm\alpha(k)}+ e^{ i\Pi} \Pi\ket{0} 
\end{equation}
which in the limit of $\Gamma^2 \gg \braket{0|\Pi^2|0}$ gives us the following constraint in (1+1) dimensions
\begin{equation}
    \left( \lambda_{\varphi} \int dk \, |\tilde{F}_{\nu}(k)|^2 \right)^2 \gg \frac{1}{2} \int dk \, \omega(k)\, |\tilde{F}_{\nu}(k)|^2. 
\end{equation} 
This constraint specifically allows for the application of $e^{ix\Pi}$ to a coherent state and results in a phase constant. Furthermore, constraining this phase constant by 
\begin{equation}\label{capital gamma constraint}
    \Gamma \coloneqq \lambda_{\Pi} \lambda_{\varphi} \int dk \, |\tilde{F}_{\nu}(k)|^2 = \frac{\pi}{4}\, \mathrm{mod}\, 2\pi
\end{equation}
results in a Bloch rotation of the output qubit A's states that eliminates the first batch of unwanted states\footnote{For a more detailed explanation of these constraints see Ref.~\cite{Simidzija2020Transmission}.}. This constraint will remain throughout the rest of this letter. 

The second group of unwanted states, and the states directly targeted for dephasing, follow from the non-orthogonal inner product discussed at the end of the previous section. The inner product has the form,
\begin{equation}\label{coherent inner product}
    |\braket{+\alpha(k)|-\alpha(k)}| = \exp{\left[-(\lambda_{\varphi})^2 \,  \int \frac{dk}{2\omega(k)} \, |\tilde{F}_{\nu}(k)|^2\right]}.
\end{equation}
Choosing a Gaussian smearing function with Gaussian width $\sigma$, and setting $\lambda \gg \sigma$ renders these terms approximately zero. With these constraints, the coherent information of the field mediated communication channel from qubit A to qubit B in Fig.~\ref{UDWQC Channel without starburst} grows toward one as we in increase the strength of the coupling. Which is demonstrated in Fig. [4] of Ref.~\cite{Simidzija2020Transmission}, as well as Figs.~\ref{Noisy_Channel_figures}.

These recent results are well understood in the RQI community, but follow from the linear nature of the field observables in Eqs.~\ref{scalar field} and \ref{conjugate momentum}. For certain nonlinear cases, it becomes a challenge to redefine the coherent amplitudes to include the coupling and smearing constants. It is for this reason, as well as previously discussed reasons, we aim to generalize these UDW gates to be understood as bosonic dephasing channels.

\section{Unruh--DeWitt Bosonic Dephasing Channels} \label{Unruh--DeWitt Dephasing Channels}
\subsection{The UDW Channel as a Bosonic Dephasing Channel} \label{The UDW Channel as a Dephasing Channel}

In quantum Shannon theory it is often useful to discuss multipartite channels that experience decoherence as dephasing channels, usually under the+ guise of interactions with the environment. To that regard, we may see a dephasing channel\footnote{It is the prerogative of the author that quantum channels including fields and qubits be denoted with $\Xi$ and channels that contain only qubits $\mathcal{N}$.} written in the Fock basis as
\begin{equation}\label{other form of dephasing channel}
    \mathcal{N}_{\gamma}(\rho) = \sum_{m,n=0}^{\infty} e^{-\frac{\gamma}{2}(n-m)} (c_nc^*_m)\ket{n}\bra{m}
\end{equation}
where $\gamma$ is the factor that determines the rate of decoherence. For example as $\gamma \rightarrow \infty$ the diagonal terms vanish. We revisit the concept of dephasing due to the environment in Sec.~\ref{Canonical Dephasing Channel}.

In case of the standard UDW channel, we aim to show that the dephasing is a necessity following the non-orthogonality of the coherent states. Generating the off-diagonals of the coherent state density matrix, scrambles the encoded information from $\hat{U}_{A\varphi}$ as the controlled structure of our unitaries encodes (decodes) QI onto (off of) the diagonals of the coherent state density matrix. Utilizing the dephasing perspective, along with our free parameters to send these states to zero, can accomplish near perfect channel capacity.

To show the relation between the current formalism and the proposed dephasing perspective, we can reformulate the state transfer channel of Ref.~\cite{Simidzija2020Transmission} to
\begin{multline} \label{decoherence two Unitary projector and field}
    \hat{U}_{\nu\varphi} = \sum_{z,x \in \pm}\hat{P}_x\hat{P}_z\otimes \exp\left[ix\int dk\,\sqrt{\gamma_{\Pi}(k)}\hat{\Pi}_{\nu}(k)\right]\\ \times\exp\left[iz\int dk\,\sqrt{\gamma_{\varphi}(k)}\hat{\varphi}_{\nu}(k)\right]
\end{multline}
where we have set
\begin{flalign}
   \hat{\varphi}(\tilde{\gamma})_{\nu} &\coloneqq \lambda_{\varphi} \int dk \tilde{F}_{\nu}(k) \hat{\varphi}(k,t_{\nu})= \int dk \sqrt{\gamma_{\varphi}(k)}\hat{\varphi}(k,t_{\nu})\label{gamma for phi}\\
    \hat{\Pi}(\tilde{\gamma})_{\nu}&\coloneqq \lambda_{\Pi} \int dk \tilde{F}_{\nu}(k) \hat{\Pi}(k,t_{\nu})= \int dk \sqrt{\gamma_{\Pi}(k)}\hat{\Pi}(k,t_{\nu})\label{gamma for pi}.
\end{flalign}
Instead of redefining the coherent amplitudes to include our free parameters as we did in Sec.~\ref{Choosing the Correct Parameters}, we have wrapped the free parameters up in some function $\gamma_{O}(k)$ and leave the coherent amplitudes as they exist in the field observables $\hat{O}$. This simplification allows for general treatments of the channel regardless of the linearity of the ladder operators in the field observables. 

Now carrying out the inner product of Eq.~\ref{coherent inner product} yields 
\begin{equation}\label{decoherence coherent inner product}
    |\braket{+\sqrt{\gamma_{\varphi}(k)}\alpha(k)|-\sqrt{\gamma_{\varphi}(k)}\alpha(k)}| = e^{-2\gamma_{\varphi}(k) |\alpha(k)|^2}
\end{equation}
where we have utilized the non-orthogonality of coherent states identity 
\begin{equation}\label{coherent state orthogonality identity}
    \braket{\beta(k)|\alpha(k)} = \exp\left(-\frac{1}{2}|\alpha(k)|^2-\frac{1}{2}|\beta(k)|^2 + \beta^*(k)\alpha(k) \right).
\end{equation}
It is obvious from Eq.\ref{decoherence coherent inner product} that the coherent states are part of a bosonic dephasing channel $\Xi_{\varphi \rightarrow \varphi'}$ with dephasing constant $\sqrt{\gamma_O(k)}$. However, unlike canonical dephasing channels the coupling, smearing, and switching are the mechanisms of dephasing. Sending our dephasing function $\gamma_{\varphi}(k) \rightarrow \infty$ for any value $k$, removes off diagonal elements from this channel. This should be expected following the constraints in Sec.~\ref{Choosing the Correct Parameters}.

\section{Applications of a UDW Channel From a Bosonic Dephasing perspective.} \label{Two Applications of a UDW Dephasing Channel}

The main result thus far has been that in the RQI channel outlined in Fig.~\ref{UDWQC Channel without starburst}, we can model the coupling constant and smearing function as a dephasing function. For the remainder of this letter, we demonstrate the results of dephasing models using numerical simulations. For this reason, we utilize a Gaussian smearing function and subsequently treat the dephasing function as a constant $\gamma_{\hat{O}}$. Since we aim to take advantage of the new perspective let's consider the canonical dephasing channel.

\subsection{Canonical Bosonic Dephasing Channel} \label{Canonical Dephasing Channel}
The canonical dephasing channel is one where the system interacts with the environment, and the multipartite quantum signal depreciates. Therefore, given the paradoxical result in our UDW system, outlining such a dephasing channel (one where the field is allowed to interact with the environment), may provide such a boost in QI through our channel. However, given that dephasing channels acts on the diagonal components of the density matrix we may find that the unwanted states remain undeterred. 

To evaluate this channel, we begin with a standard unitary that takes advantage of the number operator
\begin{equation}\label{canonical dephasing unitary}
    \hat{U}_{\mathrm{E}} = e^{-i\sqrt{\gamma_{E}}\hat{a}^{\dagger}\hat{a}(\hat{b}+\hat{b}^{\dagger})}
\end{equation}
where $\gamma_{E}$ is the dephasing constant associated with the interaction between the field and environment and $\hat{b}^{\dagger}(\hat{b})$ are the creation (annihilation) operators of the environment. We can model the channel's effect on the field by writing out the composite channel
\begin{equation}
     \Xi_{\varphi\rightarrow\varphi',E} = \Tr_E[\hat{U}_{\mathrm{E}}(\hat{\rho}_{1,\varphi}\otimes\ket{0}\bra{0})\hat{U}_{\mathrm{E}}^{\dagger}].
\end{equation}
Implanting this in the usual prescription of bosonic dephasing channels, similar to that of Eq.~\ref{bosonic dephasing channel for two}, we get
\begin{equation}\label{canonical bosonic dephasing channel for two}
    \Xi_{\varphi\rightarrow\varphi',E}(\hat{\rho}_{\varphi}) = \int_{-\infty}^{\infty} d\phi \, p(\phi)\, e^{-i\hat{a}^{\dagger}\hat{a}\phi}\hat{\rho}_{1,\varphi}e^{i\hat{a}^{\dagger}\hat{a}\phi},
\end{equation}
where $\Xi_{\varphi\rightarrow\varphi',E}(\hat{\rho}_{\varphi})$ represents the dephasing channel and the effects on our field $\varphi$, and $p(\phi)$ is a Gaussian probability density given by
\begin{equation}\label{probability density}
    p(\phi)=\sqrt{\frac{1}{2\pi\gamma_{E}}}e^{-\frac{1}{2}\frac{\phi^2}{\gamma_{E}}}.
\end{equation}
To evaluate this channel we convert to the Fock basis and continue by substituting in Eq.~\ref{density of rho 1 phi} and simplifying Eq.~\ref{canonical bosonic dephasing channel for two} which then becomes
\begin{multline}
     \Xi_{\varphi\rightarrow\varphi',E}(\hat{\rho}_{\varphi}) = \sum_{s,s' \in \pm} \int_{-\infty}^{\infty} d\phi \, p(\phi)\\
     \times e^{-i\hat{a}^{\dagger}\hat{a}\phi}\ket{s\sqrt{\gamma_{\varphi}}\alpha}\bra{s'\sqrt{\gamma_{\varphi}}\alpha} e^{i\hat{a}^{\dagger}\hat{a}\phi}\\
     =\sum_{s,s'\in \pm}\sum_{m,n} \int_{-\infty}^{\infty} d\phi \, p(\phi) e^{-\frac{1}{2}(\gamma_{\phi}|\alpha|^2(s^2+s'^2))}\\ \times\frac{(s\sqrt{\gamma_{\varphi}}\alpha)^n(s'\sqrt{\gamma_{\varphi}}\alpha^*)^m}{\sqrt{n!}\sqrt{m!}}e^{-i\hat{a}^{\dagger}\hat{a}\phi}\ket{n}\bra{m} e^{i\hat{a}^{\dagger}\hat{a}\phi}.  
\end{multline}
Evaluating the operators we obtain
\begin{multline}
     \Xi_{\varphi\rightarrow\varphi',E}(\hat{\rho}_{\varphi}) =\sum_{s,s'\in \pm}\sum_{m,n}  \int_{-\infty}^{\infty} d\phi \, p(\phi)e^{-i\phi(m-n)}\\ \times e^{-\frac{1}{2}(\gamma_{\phi}|\alpha|^2(s^2+s'^2))} \frac{(s\sqrt{\gamma_{\varphi}}\alpha)^n(s'\sqrt{\gamma_{\varphi}}\alpha^*)^m}{\sqrt{n!}\sqrt{m!}}\ket{n}\bra{m}.  
\end{multline}
Carrying out the integral yields 
\begin{multline}
   \Xi_{\varphi\rightarrow\varphi',E}(\hat{\rho}_{\varphi})=\sum_{s,s' \in \pm}\sum_{m,n} e^{-\frac{1}{2}(m-n)^2\sqrt{\gamma_E}}\\
   \times e^{-\frac{1}{2}(\gamma_{\varphi}|\alpha|^2(s^2+s'^2))}\frac{(s\sqrt{\gamma_{\varphi}}\alpha)^n(s'\sqrt{\gamma_{\varphi}}\alpha^*)^m}{\sqrt{n!}\sqrt{m!}}\ket{n}\bra{m}.
\end{multline}
It is evident that when $m=n$, we get no effect from the dephasing channel. Moreover, when considering the change that this dephasing makes to our original correlator in Eq.~\ref{Expanded out Channel}, we trace over the final state of the field and assess the inner product of our new coherent states. What we find is that non-zero results require $m=n$. Therefore, acting on the field with the number operator results in an inner product identical to Eq.~\ref{decoherence coherent inner product} and subsequently will not change the coherent information of the UDW channel. This result indicates that in this idealized dephasing channel, the environment will have no effect on the quantum information through our channel. Coincidentally, the condensed matter systems where these channels have been proposed, take advantage of topologically protected edge states to transmit quantum information \cite{aspling2022design}. Therefore, interactions with the environment were expected to be minimal regardless. 

\subsection{Cross-Talk Noise}\label{UDW Quantum Interference}
\subsubsection{Setting up a UDW Noise Channel} 
\begin{figure}[t!]
\includegraphics[width=.45\textwidth]{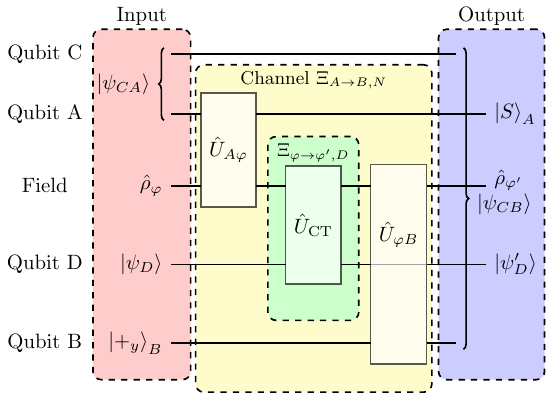}
\caption{ Introducing another UDW gate acts as additional information entering the channel and subsequently will introduce cross-talk noise to the original channel in Eq.~\ref{UDWQC_Channel}.} \label{UDWQC_Channel_Two_Encoders}
\end{figure}

One possible generation of unwanted disturbances of quantum information in proposed condensed matter approaches to UDW channels, is unwanted interactions with detectors, a type of Cross-Talk (CT) noise (see Fig.~\ref{UDWQC_Channel_Two_Encoders}). Traditional calculations of these additional interactions would increase the amount of vertex operators in the correlator of Eq.~\ref{Expanded out Channel}, making computational times significantly longer as well as more involved. However, considering the dephasing perspective, we instead can incorporate additional detector effects into the dephasing parameter.  

Let's assume that the new gate is defined as
\begin{equation}\label{single field Unitary 2}
    \hat{U}_{\mathrm{CT}} =\sum_{z \in \pm} \hat{P}_z\otimes \exp\left[iz\sqrt{\gamma_{N}}\hat{\varphi}_{\nu}\right],
\end{equation}
 similar to that of Eq.~\ref{SimpleUnitary projector and field}. Following the prescription in Sec.~\ref{Canonical Dephasing Channel} we model the channel as
\begin{equation}\label{bosonic dephasing channel for two}
    \Xi_{\varphi \rightarrow \varphi', \mathrm{N}}(\hat{\rho}_{\varphi}) = \int_{-\infty}^{\infty} d\phi \, p(\phi)\, e^{i\hat{\varphi}\phi}\hat{\rho}_{1,\varphi}e^{-i\hat{\varphi}\phi}
\end{equation}
which expresses the channel $\Xi_{\varphi \rightarrow \varphi',\mathrm{N}}(\hat{\rho}_{\varphi})$, a noisy composite channel describing the field evolution throughout the channel $\Xi_{A \rightarrow B,\mathrm{N}}$. The channel indicates the probability (which follows a random probably distribution $p(x)$ \cite{lami2023exact,Arqand2020Quantum}) that the CT detector produces a different coherent state. For this calculation we will utilize the probability distribution of Eq.~\ref{probability density} with the substitution of the dephasing parameter $\gamma_N$. $\hat{\rho}_{1,\varphi}$ is the state after the interaction between the field and qubit A and has the explicit density form
\begin{flalign}
       \hat{\rho}_{1,\varphi} &= \sum_{s,s' \in \pm} \ket{s\sqrt{\gamma_{\varphi}}\alpha}\bra{s'\sqrt{\gamma_{\varphi}}
       \alpha}\\
       &= \sum_{s,s' \in \pm} e^{s\sqrt{\gamma_{\varphi}}\hat{\varphi}}\ket{0}\bra{0}e^{-s'\sqrt{\gamma_{\varphi}}\hat{\varphi}}\label{density of rho 1 phi}.
\end{flalign}
Substituting Eq.\ref{density of rho 1 phi} into Eq.\ref{bosonic dephasing channel for two} we get
\begin{flalign}
    \Xi_{\varphi \rightarrow \varphi',\mathrm{N}}(\hat{\rho}_{\varphi}) &= \sum_{s,s'\in\pm}\int_{-\infty}^{\infty} d\phi \, p(\phi)\notag\\
    &\times e^{i(\phi+\sqrt{\gamma}_{\varphi})\hat{\varphi}}\hat{\rho}_{\varphi}e^{i(\phi+\sqrt{\gamma}_{\varphi})\hat{\varphi}}\\
    &=\sum_{s,s'\in\pm}\int_{-\infty}^{\infty} d\phi \, p(\phi) \notag\\
     & \times \ket{s(\phi+\sqrt{\gamma}_{\varphi})\alpha}\bra{s'(\phi+\sqrt{\gamma}_{\varphi})\alpha}.\label{final form of dephasing channel}
\end{flalign} 

As was shown in Sec.~\ref{UDW State Transfer Channel}, the unwanted states result from the correlator formed by the partial trace over the field and the non-orthogonality of coherent states. Tracing over $\Xi_{\varphi \rightarrow \varphi',\mathrm{N}}(\hat{\rho}_{\varphi})$ to reproduce the effect of the correlator in Eq.~\ref{Expanded out Channel} will demonstrate the changes made to the inner product of the coherent states which we have denoted by $\braket{\Xi_{\varphi \rightarrow \varphi',\mathrm{N}}(\hat{\rho}_{\varphi})}$ and defined as
\begin{multline}\label{EV of dephasing channel 2}
    \Tr \Xi_{\varphi \rightarrow \varphi',\mathrm{N}}(\hat{\rho}_{\varphi})\equiv \braket{\Xi_{\varphi \rightarrow \varphi',\mathrm{N}}(\hat{\rho}_{\varphi})} = \sum_{s,s' \in \pm} \int_{-\infty}^{\infty} d\phi \, p(\phi)\\
    \times \braket{s'(\phi+\sqrt{\gamma}_{\varphi})\alpha|s(\phi+\sqrt{\gamma}_{\varphi})\alpha}\\
    = \sum_{s,s' \in \pm} \int_{-\infty}^{\infty} d\phi \, p(\phi)\, \exp \bigg(-\frac{1}{2} s'^2(\phi+\sqrt{\gamma}_{\varphi})^2|\alpha|^2 \\
    -\frac{1}{2} s^2(\phi+\sqrt{\gamma}_{\varphi})^2|\alpha|^2+ s's(\phi+\sqrt{\gamma}_{\varphi})^2|\alpha|^2\bigg)\\
    =\sum_{s,s' \in \pm} \int_{-\infty}^{\infty} d\phi \, p(\phi)\, \exp \bigg( -\frac{1}{2}  (\phi+\sqrt{\gamma}_{\varphi})^2|\alpha|^2(s'-s)^2\bigg)
\end{multline}
where we have utilized Eq.~\ref{coherent state orthogonality identity} as we did in Sec.~\ref{The UDW Channel as a Dephasing Channel}. Notice Eq.~\ref{EV of dephasing channel 2} has the same structure as Eq.~\ref{decoherence coherent inner product} but allows new dephasing construction of the parameters. As expected when $s'=s$ the expectation value is trivially one and will not effect the outcome of channel Eq.\ref{UDWQC_Channel}. However, when $s' \neq s$ we now have two parameters, $\gamma_{\varphi}$ and $\phi$, to enforce dephasing of unwanted states. 

\begin{figure*}[t]
    \centering
    \begin{subfigure}{0.45\textwidth}
        \includegraphics[width=\textwidth]{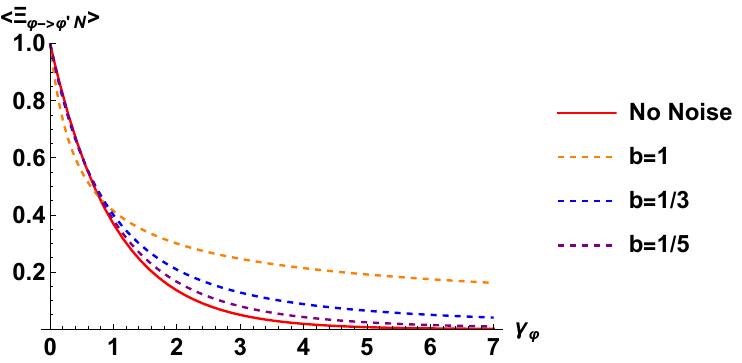}
        \caption{As $ b \rightarrow 0$ the coupling as well as the noise is turning off.}\label{Noisy_LT1_Decoherence_Plots}
    \end{subfigure} \hfill
    \begin{subfigure}{0.45\textwidth}
        \includegraphics[width=\textwidth]{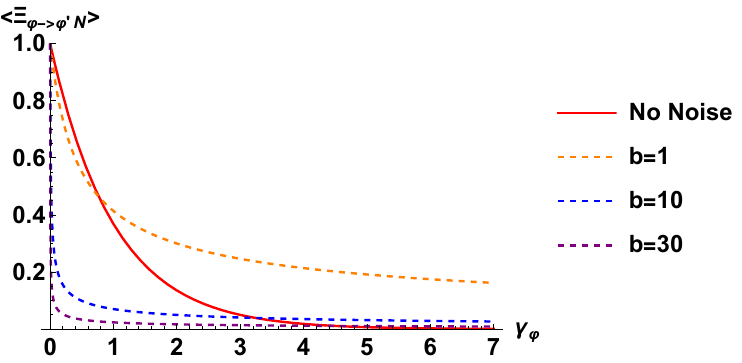}
        \caption{As $ b \rightarrow \infty$ it takes over the dephasing process for smaller values of $\gamma_{\varphi}$}\label{Noisy_GT1_Decoherence_Plots}
    \end{subfigure}
    
    \begin{subfigure}{0.45\textwidth}
        \includegraphics[width=\textwidth]{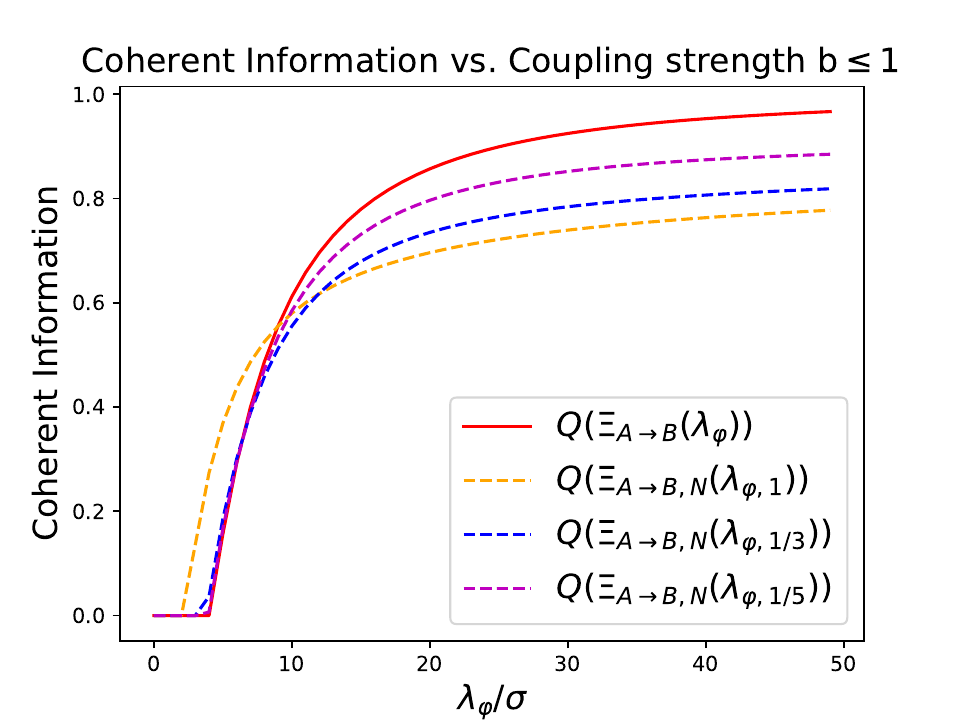}
        \caption{As the noise is turning off the quantum capacity is increasing.}\label{UDW_Channel_Capacity_Noisy_LT1_Decoherence}
    \end{subfigure} \hfill
    \begin{subfigure}{0.45\textwidth}
        \includegraphics[width=\textwidth]{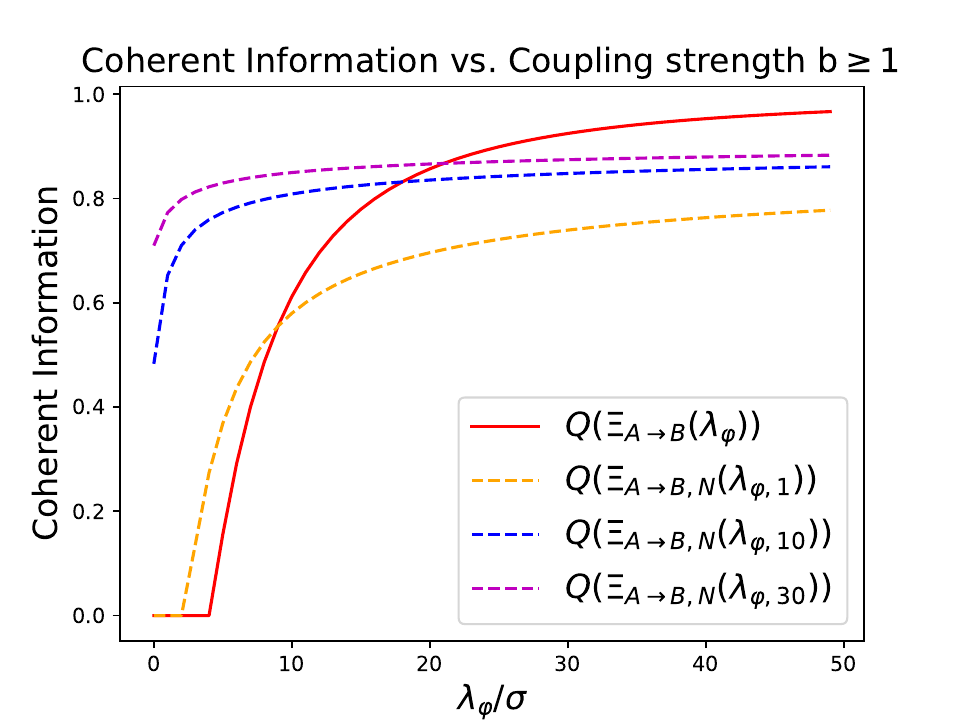}
        \caption{For large $b$ and small values of $\lambda_{\varphi}$ the additional interaction can result in better quantum capacity.}\label{UDW_Channel_Capacity_Noisy_GT1_Decoherence}
    \end{subfigure}
\caption{Comparing the coherent information of channels with and without noise, it is clear to see that the noisy channel $\Xi_{A\rightarrow B,N}(\lambda_{\varphi,b})$ reaches a channel capacity of one at a slower rate. At $b=1$ the noisy signal is the highest, this is due to a signal boost in the unwanted states. Regardless, we can see in (d) an increase in the lower bound of coherent information for small values of $\gamma_{\varphi}$.} \label{Noisy_Channel_figures}
\end{figure*}

Plugging the above probability distribution into Eq.~\ref{EV of dephasing channel 2} and evaluating the integral for $s'\neq s$ we get
\begin{equation} \label{nosiy channel final inner product}
     \braket{\Xi_{\varphi \rightarrow \varphi',\mathrm{N}}(\hat{\rho}_{\varphi})} = \frac{\exp{\bigg(-\frac{2 \gamma_{\varphi} |\alpha|^2}{1+ 4|\alpha|^2b^2\gamma_{\varphi}}\bigg)}}{\sqrt{1+ 4|\alpha|^2b^2\gamma_{\varphi}}}
\end{equation}
where we have redefined our CT dephasing parameter as a multiple of the original $\gamma_N=b\gamma_{\varphi}$. Eq.~\ref{nosiy channel final inner product} is directly comparable to Eq.~\ref{decoherence coherent inner product}. It is straightforward to see in Fig.~\ref{Noisy_LT1_Decoherence_Plots} as $b \rightarrow 0$ the interaction with the noise is turned off, at $b=1$ the channel is the noisiest, and as $b\rightarrow \infty$ $\gamma_N$ acts as the primary dephasing factor for small values of $\gamma_{\varphi}$ which is demonstrated in Fig.~\ref{Noisy_GT1_Decoherence_Plots}.

\subsubsection{Noisy UDW Effects on Coherent Information}
To understand the difference these new parameters make to the coherent information we can look at the definition of $\gamma_{\varphi}$ in Eq.~\ref{gamma for phi}. Since $\gamma_{\varphi}$ is a function of $\lambda_{\varphi}$ we can set up the new inner product in terms of a new coupling constant $\lambda_{\varphi,b}$ given by the relation
\begin{multline}
     \frac{\exp{\bigg(-\frac{2\gamma_{\varphi} |\alpha|^2}{1+4b^2\gamma_{\varphi}|\alpha|^2}\bigg)}}{\sqrt{1+4b^2\gamma_{\varphi}|\alpha|^2}} = \exp\bigg[-2(\lambda_{\varphi,b})^2\\ \times \int dk \, |\tilde{F}_{\nu}(k)|^2 |\alpha|^2\bigg].
\end{multline}

We then solve for $\lambda_{\varphi,b}$ in terms of $\lambda_{\varphi}$ and ascertain
\begin{multline}
    \lambda_{\varphi,b}=\bigg[ \frac{\lambda_{\varphi}^2}{1+\frac{4b^2|\alpha|^2\lambda_{\varphi}^2}{\sqrt{(2\pi)^3}\sigma}} -\frac{\sqrt{(2\pi)^3}\sigma}{2|\alpha|^2}\\ \times\ln\left(1+\frac{4b^2|\alpha|^2\lambda_{\varphi}^2}{\sqrt{(2\pi)^3}\sigma}\right)\bigg]^{\frac{1}{2}}
\end{multline}
where the term $\sqrt{(2\pi)^3} \sigma$ is a consequence of Gaussian smearing with width $\sigma$. Calculating channel capacity with this new value $\lambda_{\varphi,b}$, while keeping the remaining parameters intact, still reaches near perfect channel capacity but reaches it slower, when $b\leq 1$ demonstrated by Fig.~\ref{UDW_Channel_Capacity_Noisy_LT1_Decoherence}. Furthermore, for high values of $b$ and low values of $\lambda_{\varphi}$ one would expect $b$ to act as the primary dephasing factor as shown in Fig.~\ref{UDW_Channel_Capacity_Noisy_GT1_Decoherence}.

\section{Summary and Results}
We have shown that the formalism of quantum channels produced by UDW detectors provides a bosonic dephasing channel perspective. With this perspective, we have demonstrated that the purpose of the dephasing is to remove unwanted states that lead to the scrambling of the quantum information encoded on the field. We have shown that the dephasing constant can be written in terms of the strength of coupling between the qubit and field.

We aimed to demonstrate several applications of the dephasing perspective that provide interpretations of unwanted interactions. Firstly, we applied the canonical form of a bosonic dephasing channel, allowing the system to interact with the environment. In this idealized dephasing channel, the prior constraints on the off-diagonals were the only effects that remained when tracing over the field. Subsequently, this indicated no additional noise generated from interacting with the environment under this prescription.

Secondly, how does coherent information change given some noise due to additional UDW detectors? To evaluate this numerically, we presented a unitary operation that introduces CT noise to the system and calculated how that noise affects the strength of coupling between the qubits and fields. Figure~\ref{Noisy_Channel_figures} demonstrates how the additional noise can affect the propagation of quantum information through the channel given in Eq.~\ref{UDWQC_Channel}. These values make intuitive sense as one may expect an additional signal boost to non-orthogonal off-diagonal elements of the density matrix in Eq.~\ref{final form of dephasing channel} which act to scramble the QI.

An open problem remaining is the possibility of writing down a unitary dephasing channel that increases coherent information overall. One might notice that non-unitary interactions that increase the strength of the diagonal elements of the coherent state density matrix while decreasing the off-diagonal elements are possible. However, given the nature of dephasing channels, accomplishing this with unitary operations requires much care. 

\section{Acknowledgments}
We thank Ludovico Lami and Mark Wilde for useful discussions on dephasing channels and key insights into the role coherent states play in quantum information theory. We would also like to thank Justin Kulp for several conversations, resulting in more precise and quicker data collection and interpretation. 

\bibliography{UDW_Dephasing_bib}

\begin{thebibliography}{16}%
\makeatletter
\providecommand \@ifxundefined [1]{%
 \@ifx{#1\undefined}
}%
\providecommand \@ifnum [1]{%
 \ifnum #1\expandafter \@firstoftwo
 \else \expandafter \@secondoftwo
 \fi
}%
\providecommand \@ifx [1]{%
 \ifx #1\expandafter \@firstoftwo
 \else \expandafter \@secondoftwo
 \fi
}%
\providecommand \natexlab [1]{#1}%
\providecommand \enquote  [1]{``#1''}%
\providecommand \bibnamefont  [1]{#1}%
\providecommand \bibfnamefont [1]{#1}%
\providecommand \citenamefont [1]{#1}%
\providecommand \href@noop [0]{\@secondoftwo}%
\providecommand \href [0]{\begingroup \@sanitize@url \@href}%
\providecommand \@href[1]{\@@startlink{#1}\@@href}%
\providecommand \@@href[1]{\endgroup#1\@@endlink}%
\providecommand \@sanitize@url [0]{\catcode `\\12\catcode `\$12\catcode
  `\&12\catcode `\#12\catcode `\^12\catcode `\_12\catcode `\%12\relax}%
\providecommand \@@startlink[1]{}%
\providecommand \@@endlink[0]{}%
\providecommand \url  [0]{\begingroup\@sanitize@url \@url }%
\providecommand \@url [1]{\endgroup\@href {#1}{\urlprefix }}%
\providecommand \urlprefix  [0]{URL }%
\providecommand \Eprint [0]{\href }%
\providecommand \doibase [0]{https://doi.org/}%
\providecommand \selectlanguage [0]{\@gobble}%
\providecommand \bibinfo  [0]{\@secondoftwo}%
\providecommand \bibfield  [0]{\@secondoftwo}%
\providecommand \translation [1]{[#1]}%
\providecommand \BibitemOpen [0]{}%
\providecommand \bibitemStop [0]{}%
\providecommand \bibitemNoStop [0]{.\EOS\space}%
\providecommand \EOS [0]{\spacefactor3000\relax}%
\providecommand \BibitemShut  [1]{\csname bibitem#1\endcsname}%
\let\auto@bib@innerbib\@empty
\bibitem [{\citenamefont
  {Martin-Martinez}(2011)}]{Martin-Martinez2011relativistic}%
  \BibitemOpen
  \bibfield  {author} {\bibinfo {author} {\bibfnamefont {E.}~\bibnamefont
  {Martin-Martinez}},\ }\emph {\bibinfo {title} {{Relativistic Quantum
  Information: developments in Quantum Information in general relativistic
  scenarios}}},\ \href@noop {} {Ph.D. thesis},\ \bibinfo  {school} {Waterloo
  U.} (\bibinfo {year} {2011}),\ \Eprint {https://arxiv.org/abs/1106.0280}
  {arXiv:1106.0280 [quant-ph]} \BibitemShut {NoStop}%
\bibitem [{\citenamefont {Simidzija}\ \emph {et~al.}(2020)\citenamefont
  {Simidzija}, \citenamefont {Ahmadzadegan}, \citenamefont {Kempf},\ and\
  \citenamefont {Mart\'{\i}n-Mart\'{\i}nez}}]{Simidzija2020Transmission}%
  \BibitemOpen
  \bibfield  {author} {\bibinfo {author} {\bibfnamefont {P.}~\bibnamefont
  {Simidzija}}, \bibinfo {author} {\bibfnamefont {A.}~\bibnamefont
  {Ahmadzadegan}}, \bibinfo {author} {\bibfnamefont {A.}~\bibnamefont
  {Kempf}},\ and\ \bibinfo {author} {\bibfnamefont {E.}~\bibnamefont
  {Mart\'{\i}n-Mart\'{\i}nez}},\ }\bibfield  {title} {\bibinfo {title}
  {Transmission of quantum information through quantum fields},\ }\href
  {https://doi.org/10.1103/PhysRevD.101.036014} {\bibfield  {journal} {\bibinfo
   {journal} {Phys. Rev. D}\ }\textbf {\bibinfo {volume} {101}},\ \bibinfo
  {pages} {036014} (\bibinfo {year} {2020})}\BibitemShut {NoStop}%
\bibitem [{\citenamefont {Sachs}\ \emph {et~al.}(2017)\citenamefont {Sachs},
  \citenamefont {Mann},\ and\ \citenamefont
  {Martin-Martinez}}]{Sachs2017Entanglement}%
  \BibitemOpen
  \bibfield  {author} {\bibinfo {author} {\bibfnamefont {A.}~\bibnamefont
  {Sachs}}, \bibinfo {author} {\bibfnamefont {R.~B.}\ \bibnamefont {Mann}},\
  and\ \bibinfo {author} {\bibfnamefont {E.}~\bibnamefont {Martin-Martinez}},\
  }\bibfield  {title} {\bibinfo {title} {{Entanglement harvesting and
  divergences in quadratic Unruh-DeWitt detector pairs}},\ }\href
  {https://doi.org/10.1103/PhysRevD.96.085012} {\bibfield  {journal} {\bibinfo
  {journal} {Phys. Rev. D}\ }\textbf {\bibinfo {volume} {96}},\ \bibinfo
  {pages} {085012} (\bibinfo {year} {2017})},\ \Eprint
  {https://arxiv.org/abs/1704.08263} {arXiv:1704.08263 [quant-ph]} \BibitemShut
  {NoStop}%
\bibitem [{\citenamefont {Simidzija}\ \emph {et~al.}(2018)\citenamefont
  {Simidzija}, \citenamefont {Jonsson},\ and\ \citenamefont
  {Mart\'{\i}n-Mart\'{\i}nez}}]{Simidzija2018General}%
  \BibitemOpen
  \bibfield  {author} {\bibinfo {author} {\bibfnamefont {P.}~\bibnamefont
  {Simidzija}}, \bibinfo {author} {\bibfnamefont {R.~H.}\ \bibnamefont
  {Jonsson}},\ and\ \bibinfo {author} {\bibfnamefont {E.}~\bibnamefont
  {Mart\'{\i}n-Mart\'{\i}nez}},\ }\bibfield  {title} {\bibinfo {title} {General
  no-go theorem for entanglement extraction},\ }\href
  {https://doi.org/10.1103/PhysRevD.97.125002} {\bibfield  {journal} {\bibinfo
  {journal} {Phys. Rev. D}\ }\textbf {\bibinfo {volume} {97}},\ \bibinfo
  {pages} {125002} (\bibinfo {year} {2018})}\BibitemShut {NoStop}%
\bibitem [{\citenamefont {Yamaguchi}\ and\ \citenamefont
  {Kempf}(2023)}]{Yamaguchi2023Entanglement}%
  \BibitemOpen
  \bibfield  {author} {\bibinfo {author} {\bibfnamefont {K.}~\bibnamefont
  {Yamaguchi}}\ and\ \bibinfo {author} {\bibfnamefont {A.}~\bibnamefont
  {Kempf}},\ }\bibfield  {title} {\bibinfo {title} {{Entanglement is better
  teleported than transmitted}},\ }\href@noop {} {\  (\bibinfo {year}
  {2023})},\ \Eprint {https://arxiv.org/abs/2301.13212} {arXiv:2301.13212
  [quant-ph]} \BibitemShut {NoStop}%
\bibitem [{\citenamefont {Tjoa}\ and\ \citenamefont
  {Gallock-Yoshimura}(2022)}]{Tjoa2022Channel}%
  \BibitemOpen
  \bibfield  {author} {\bibinfo {author} {\bibfnamefont {E.}~\bibnamefont
  {Tjoa}}\ and\ \bibinfo {author} {\bibfnamefont {K.}~\bibnamefont
  {Gallock-Yoshimura}},\ }\bibfield  {title} {\bibinfo {title} {Channel
  capacity of relativistic quantum communication with rapid interaction},\
  }\bibfield  {journal} {\bibinfo  {journal} {Physical Review D}\ }\textbf
  {\bibinfo {volume} {105}},\ \href
  {https://doi.org/10.1103/physrevd.105.085011} {10.1103/physrevd.105.085011}
  (\bibinfo {year} {2022})\BibitemShut {NoStop}%
\bibitem [{\citenamefont {Aspling}\ \emph {et~al.}(2022)\citenamefont
  {Aspling}, \citenamefont {Marohn},\ and\ \citenamefont
  {Lawler}}]{aspling2022design}%
  \BibitemOpen
  \bibfield  {author} {\bibinfo {author} {\bibfnamefont {E.~W.}\ \bibnamefont
  {Aspling}}, \bibinfo {author} {\bibfnamefont {J.~A.}\ \bibnamefont
  {Marohn}},\ and\ \bibinfo {author} {\bibfnamefont {M.~J.}\ \bibnamefont
  {Lawler}},\ }\href@noop {} {\bibinfo {title} {Design constraints for
  unruh-dewitt quantum computers}} (\bibinfo {year} {2022}),\ \Eprint
  {https://arxiv.org/abs/2210.12552} {arXiv:2210.12552 [quant-ph]} \BibitemShut
  {NoStop}%
\bibitem [{\citenamefont {Wilde}()}]{Wilde2011From}%
  \BibitemOpen
  \bibfield  {author} {\bibinfo {author} {\bibfnamefont {M.~M.}\ \bibnamefont
  {Wilde}},\ }\bibfield  {title} {\bibinfo {title} {Preface to the second
  edition},\ }\href {https://doi.org/10.1017/9781316809976.001} {\bibinfo
  {journal} {Quantum Information Theory}\ ,\ \bibinfo {pages}
  {xi–xii}}\BibitemShut {NoStop}%
\bibitem [{\citenamefont {Lami}\ and\ \citenamefont
  {Wilde}(2023)}]{lami2023exact}%
  \BibitemOpen
\bibfield  {journal} {  }\bibfield  {author} {\bibinfo {author} {\bibfnamefont
  {L.}~\bibnamefont {Lami}}\ and\ \bibinfo {author} {\bibfnamefont {M.~M.}\
  \bibnamefont {Wilde}},\ }\bibfield  {title} {\bibinfo {title} {Exact solution
  for the quantum and private capacities of bosonic dephasing channels},\
  }\href@noop {} {\bibfield  {journal} {\bibinfo  {journal} {Nature Photonics}\
  ,\ \bibinfo {pages} {1}} (\bibinfo {year} {2023})}\BibitemShut {NoStop}%
\bibitem [{\citenamefont {Arqand}\ \emph {et~al.}(2020)\citenamefont {Arqand},
  \citenamefont {Memarzadeh},\ and\ \citenamefont
  {Mancini}}]{Arqand2020Quantum}%
  \BibitemOpen
  \bibfield  {author} {\bibinfo {author} {\bibfnamefont {A.}~\bibnamefont
  {Arqand}}, \bibinfo {author} {\bibfnamefont {L.}~\bibnamefont {Memarzadeh}},\
  and\ \bibinfo {author} {\bibfnamefont {S.}~\bibnamefont {Mancini}},\
  }\bibfield  {title} {\bibinfo {title} {Quantum capacity of a bosonic
  dephasing channel},\ }\bibfield  {journal} {\bibinfo  {journal} {Physical
  Review A}\ }\textbf {\bibinfo {volume} {102}},\ \href
  {https://doi.org/10.1103/physreva.102.042413} {10.1103/physreva.102.042413}
  (\bibinfo {year} {2020})\BibitemShut {NoStop}%
\bibitem [{\citenamefont {Dehdashti}\ \emph {et~al.}(2022)\citenamefont
  {Dehdashti}, \citenamefont {Notzel},\ and\ \citenamefont {van
  Loock}}]{dehdashti2022quantum}%
  \BibitemOpen
  \bibfield  {author} {\bibinfo {author} {\bibfnamefont {S.}~\bibnamefont
  {Dehdashti}}, \bibinfo {author} {\bibfnamefont {J.}~\bibnamefont {Notzel}},\
  and\ \bibinfo {author} {\bibfnamefont {P.}~\bibnamefont {van Loock}},\
  }\href@noop {} {\bibinfo {title} {Quantum capacity of a deformed bosonic
  dephasing channel}} (\bibinfo {year} {2022}),\ \Eprint
  {https://arxiv.org/abs/2211.09012} {arXiv:2211.09012 [quant-ph]} \BibitemShut
  {NoStop}%
\bibitem [{\citenamefont {zhen Jiang}\ and\ \citenamefont
  {yu~Chen}(2010)}]{Jiang2010Evaluating}%
  \BibitemOpen
  \bibfield  {author} {\bibinfo {author} {\bibfnamefont {L.}~\bibnamefont {zhen
  Jiang}}\ and\ \bibinfo {author} {\bibfnamefont {X.}~\bibnamefont {yu~Chen}},\
  }\bibfield  {title} {\bibinfo {title} {{Evaluating the quantum capacity of
  bosonic dephasing channel}},\ }in\ \href {https://doi.org/10.1117/12.870179}
  {\emph {\bibinfo {booktitle} {Quantum and Nonlinear Optics}}},\ Vol.\
  \bibinfo {volume} {7846},\ \bibinfo {editor} {edited by\ \bibinfo {editor}
  {\bibfnamefont {Q.}~\bibnamefont {Gong}}, \bibinfo {editor} {\bibfnamefont
  {G.-C.}\ \bibnamefont {Guo}},\ and\ \bibinfo {editor} {\bibfnamefont {Y.-R.}\
  \bibnamefont {Shen}}},\ \bibinfo {organization} {International Society for
  Optics and Photonics}\ (\bibinfo  {publisher} {SPIE},\ \bibinfo {year}
  {2010})\ p.\ \bibinfo {pages} {784613}\BibitemShut {NoStop}%
\bibitem [{\citenamefont {Leviant}\ \emph {et~al.}(2022)\citenamefont
  {Leviant}, \citenamefont {Xu}, \citenamefont {Jiang},\ and\ \citenamefont
  {Rosenblum}}]{Leviant2022quantum}%
  \BibitemOpen
  \bibfield  {author} {\bibinfo {author} {\bibfnamefont {P.}~\bibnamefont
  {Leviant}}, \bibinfo {author} {\bibfnamefont {Q.}~\bibnamefont {Xu}},
  \bibinfo {author} {\bibfnamefont {L.}~\bibnamefont {Jiang}},\ and\ \bibinfo
  {author} {\bibfnamefont {S.}~\bibnamefont {Rosenblum}},\ }\bibfield  {title}
  {\bibinfo {title} {Quantum capacity and codes for the bosonic loss-dephasing
  channel},\ }\href {https://doi.org/10.22331/q-2022-09-29-821} {\bibfield
  {journal} {\bibinfo  {journal} {Quantum}\ }\textbf {\bibinfo {volume} {6}},\
  \bibinfo {pages} {821} (\bibinfo {year} {2022})}\BibitemShut {NoStop}%
\bibitem [{\citenamefont {Kam}\ \emph {et~al.}(2023)\citenamefont {Kam},
  \citenamefont {Zhang},\ and\ \citenamefont {Feng}}]{kam2023coherent}%
  \BibitemOpen
  \bibfield  {author} {\bibinfo {author} {\bibfnamefont {C.-F.}\ \bibnamefont
  {Kam}}, \bibinfo {author} {\bibfnamefont {W.-M.}\ \bibnamefont {Zhang}},\
  and\ \bibinfo {author} {\bibfnamefont {D.-H.}\ \bibnamefont {Feng}},\
  }\href@noop {} {\emph {\bibinfo {title} {Coherent States: New Insights into
  Quantum Mechanics with Applications}}},\ Vol.\ \bibinfo {volume} {1011}\
  (\bibinfo  {publisher} {Springer Nature},\ \bibinfo {year}
  {2023})\BibitemShut {NoStop}%
\bibitem [{\citenamefont {Simidzija}\ and\ \citenamefont
  {Martín-Martínez}(2017)}]{Simidzija2017Nonperturbative}%
  \BibitemOpen
  \bibfield  {author} {\bibinfo {author} {\bibfnamefont {P.}~\bibnamefont
  {Simidzija}}\ and\ \bibinfo {author} {\bibfnamefont {E.}~\bibnamefont
  {Martín-Martínez}},\ }\bibfield  {title} {\bibinfo {title} {Nonperturbative
  analysis of entanglement harvesting from coherent field states},\ }\bibfield
  {journal} {\bibinfo  {journal} {Physical Review D}\ }\textbf {\bibinfo
  {volume} {96}},\ \href {https://doi.org/10.1103/physrevd.96.065008}
  {10.1103/physrevd.96.065008} (\bibinfo {year} {2017})\BibitemShut {NoStop}%
\bibitem [{\citenamefont {Nielsen}\ and\ \citenamefont
  {Chuang}(2010)}]{nielsen2010Quantum}%
  \BibitemOpen
  \bibfield  {author} {\bibinfo {author} {\bibfnamefont {M.~A.}\ \bibnamefont
  {Nielsen}}\ and\ \bibinfo {author} {\bibfnamefont {I.~L.}\ \bibnamefont
  {Chuang}},\ }\href {https://doi.org/10.1017/CBO9780511976667} {\emph
  {\bibinfo {title} {Quantum Computation and Quantum Information: 10th
  Anniversary Edition}}}\ (\bibinfo  {publisher} {Cambridge University Press},\
  \bibinfo {year} {2010})\BibitemShut {NoStop}%
\end{thebibliography}%
\end{document}